\documentclass[letter]{aa} 
\usepackage{graphicx}
\usepackage{txfonts,textcomp}
\modulolinenumbers[0]
\linenumbers
\renewcommand\linenumberfont{\normalfont\relax} 
\let\makeLineNumber\relax 

\usepackage[colorlinks=true, allcolors=blue]{hyperref}
\newcommand{\Msun}{\ensuremath{\mathrm{M}_\odot}}
\newcommand{\Mbh}{\ensuremath{M_\mathrm{BH}}}

\newcommand{\Lsun}{\ensuremath{\mathrm{L}_\odot}}

\newcommand{\ml}{\ensuremath{M/L}}

\newcommand{\kms}{km~s$^{-1}$}

\mathchardef\mhyphen="2D

\begin{document} 
     
    \title{Supermassive black hole mass measurement in the spiral galaxy NGC~4736}
    \subtitle{Using JWST/NIRSpec stellar kinematics}

    \author {Dieu D.\ Nguyen \inst{1}
         \and Hai N.\  Ngo \inst{2}
         \and Tinh Q.\ T.\ Le \inst{3, 1}
         \and Alister W.\ Graham \inst{4}
         \and Roberto Soria \inst{5,6} 
         \and Igor V.\ Chilingarian \inst{7,8}
         \and Niranjan Thatte \inst{9}   
         \and N.\ T.\ Phuong \inst{3, 1}
         \and Thiem Hoang \inst{10, 11}
         \and Miguel Pereira-Santaella \inst{12}
         \and Mark Durre \inst{4}
         \and Diep N.\ Pham \inst{3}  
         \and Le Ngoc Tram \inst{13}
         \and Nguyen B.\ Ngoc \inst{3}  
         \and Ng\^an~L\^e \inst{1}  
           }       
\institute{International Centre for Interdisciplinary Science and Education 07 Science Avenue, Ghenh Rang, 55121 Quy Nhon, Vietnam; Email: nddieuphys@gmail.com
              \and Faculty of Physics -- Engineering Physics, University of Science, Vietnam National University, Ho Chi Minh City, Vietnam
               \and Vietnam National Space Center, Vietnam Academy of Science and Technology, 18 Hoang Quoc Viet, Cau Giay, Hanoi, Vietnam     
               \and Centre for Astrophysics and Supercomputing, Swinburne University of Technology, Hawthorn, VIC 3122, Australia 
               \and INAF-Osservatorio Astrofisico di Torino, Strada Osservatorio 20, I-10025 Pino Torinese, Italy 
               \and Sydney Institute for Astronomy, School of Physics A28, The University of Sydney, Sydney, NSW 2006, Australia
              \and Center for Astrophysics – Harvard and Smithsonian, 60 Garden St. Cambridge, MA, 02138 USA
              \and Sternberg Astronomical Institute, M.V. Lomonosov Moscow State University, 13 Universitetsky prospect, 119992 Moscow, Russia
               \and Sub-Department of Astrophysics, Department of Physics, University of Oxford, DWB, Keble Road, Oxford OX1 3RH, UK   
               \and Korea Astronomy and Space Science Institute, Daejeon 34055, Republic of Korea
               \and Department of Astronomy and Space Science, UST, 217 Gajeong-ro, Yuseong-gu, Daejeon 34113, Republic of Korea   
               \and Instituto de F\'isica Fundamental, CSIC, Calle Serrano 123, 28006 Madrid, Spain         
               \and Leiden Observatory, Leiden University, PO Box 9513, 2300 RA Leiden, The Netherlands}           

 \date{Received 31 March 2025; accepted 12 May 2025}

\abstract 
{\small We present accurate mass measurements of the central supermassive black hole (SMBH) in NGC~4736 (M~94).\ We used the ``gold-standard'' stellar absorption features (CO band heads) at $\sim$2.3\,${\rm \mu m}$, as opposed to gas emission lines, to trace the dynamics in the nuclear region, easily resolving the SMBH's sphere of influence.         The analysis uses observations made with the integral field unit of the Near-Infrared Spectrograph (NIRSpec) on the {\it James Webb} Space Telescope and a surface brightness profile derived from {\it Hubble} Space Telescope archival images.         We used Jeans anisotropic models within a Bayesian framework, and comprehensive Markov chain Monte Carlo optimization, to determine the best-fit black hole mass, orbital anisotropy, mass-to-light ratio, and nucleus kinematical inclination. We obtained a SMBH mass $M_{\rm BH}=(1.60\pm0.16)\times10^7$ \Msun\ (1$\sigma$ random error), which is consistent with the $M_{\rm BH}$--$\sigma$ and $M_{\rm BH}$--$M_\star$ relations.        This is the first dynamical measurement of a $M_{\rm BH}$ in NGC 4736 based on the stellar kinematics observed with NIRSpec.      We thus settle a longstanding inconsistency between estimates based on nuclear emission-line tracers and the $M_{\rm BH}$--$\sigma$ relation.      Our analysis shows that NIRSpec can detect SMBHs with $M_{\rm BH,min}\approx 5\times10^6$ \Msun\  in galaxies within 5 Mpc and $\sigma\approx100$ km s$^{-1}$.} 


   \maketitle

\section{Introduction} \label{intro}

As a compact, powerful engine at a galaxy's center, supermassive black holes (SMBHs) play a crucial role in driving the evolution of their host galaxies \citep{Kormendy13}. Accurately measuring SMBH masses (\Mbh) is essential for constraining galaxy–black hole correlations and distinguishing their various coevolution pathways \citep{Graham_Sahu23}.   

The \Mbh\ estimate for NGC 4736 (M 94) has long been debated and varies significantly depending on the method and assumed distance (4.3–6 Mpc). The  \citet{Graham23a} \Mbh--$\sigma$ relation predicts \Mbh\ $\approx (1.05 \pm 0.64) \times 10^7$~\Msun\  based on a bulge stellar velocity dispersion of $\sigma = 110 \pm 5$~\kms\ \citep{Barth02}. In contrast, {\it Hubble} Space Telescope (HST) observations of the broad H$\alpha$ emission component yielded \Mbh\ $\approx 3 \times 10^4$~\Msun\ \citep{Greene05}. Mid-infrared correlations based on [\ion{Ne}{v}] and [\ion{O}{iv}] emission lines suggested \Mbh\ $=(2.3 \pm 0.4) \times 10^5$~\Msun\ and $(1.7 \pm 0.5) \times 10^5$~\Msun\ \citep{Dasyra08}, respectively.  Furthermore, the X-ray luminosity (2--10 keV band) of the point-like source spatially coincident with the nucleus is $\log_{10}L_X = 38.48$ erg s$^{-1}$ \citep{Eracleous02, Pellegrini02, vanOers17}. Together with the 5 GHz radio luminosity of the unresolved radio nucleus \citep[$\log_{10}L_R = 35.51$ erg s$^{-1}$;][]{vanOers17}, the \citet{Merloni03} fundamental plane (FP) relation estimates \Mbh\ $=8.2^{+2.1}_{-1.8}\times10^5$~\Msun.  However, alternative versions of the FP relation more suitable to the low-luminosity active galactic nucleus (LLAGN) regime predict higher masses: \Mbh\ $=1.1^{+1.2}_{-0.7}\times10^7$~\Msun\ according to \citet{Plotkin12}, and \Mbh\ $=1.6^{+2.4}_{-1.0}\times10^7$~\Msun\ according to \citet{Gultekin19}.

\begin{figure*}[!h]
\centering
        \includegraphics[width=0.68\textwidth]{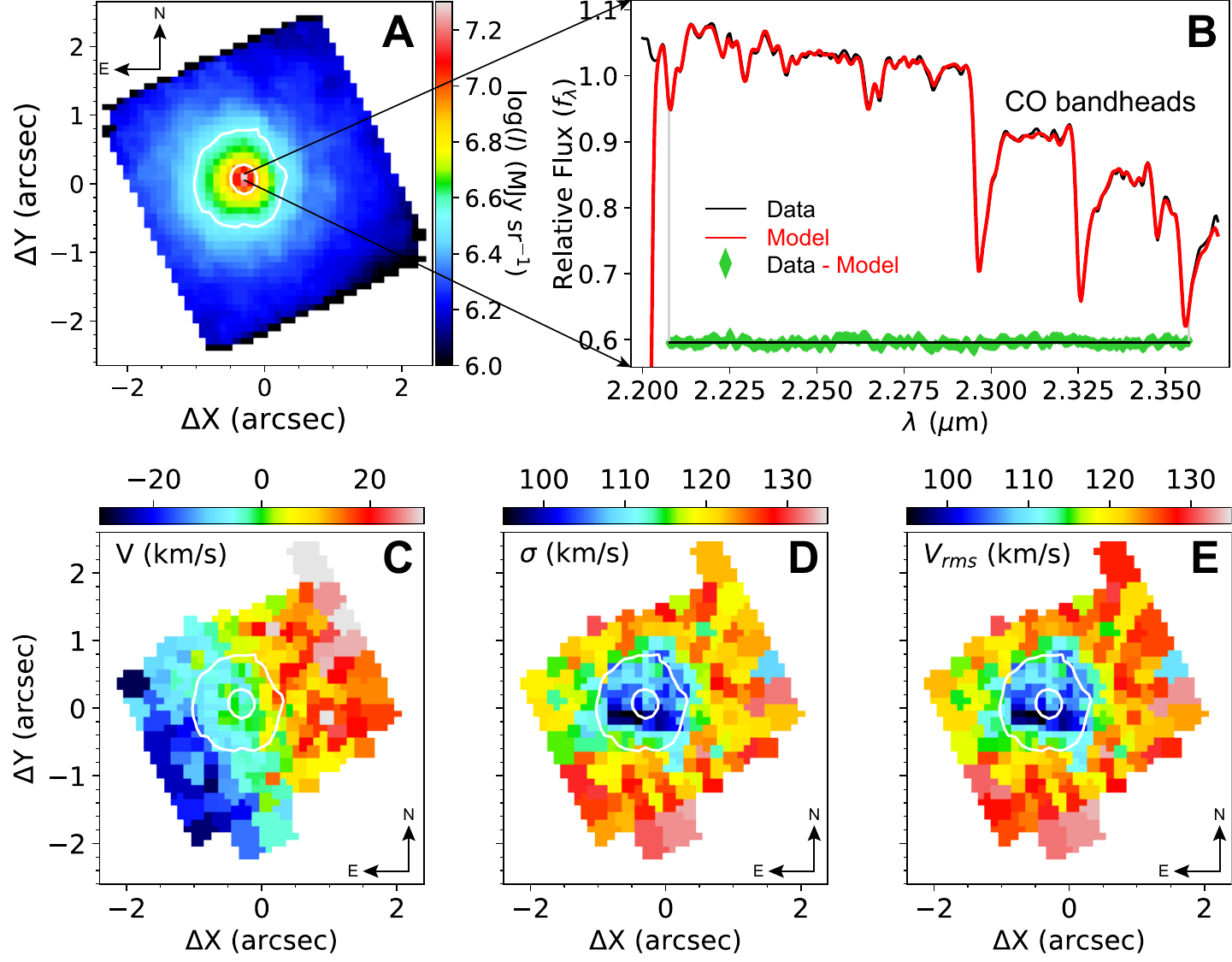} 
        \caption{\label{jwst_kine_ppxf}Stellar kinematic measurements extracted from the JWST/NIRSpec~G235H/F170LP~IFU. Panel A: Logarithmic integrated intensity collapsed along the spectral dimension (excluding the detector gap at $\sim$2.41--2.49\,${\rm \mu m}$).\ Panel B:  \textsc{pPXF} fit for the central \textsc{Voronoi} bin. Part of the central bin spectrum, which contains the stellar absorption features at $\sim$2.3\,${\rm \mu m}$ (CO band heads), is plotted in black, and the best-fit empirical X-shooter stellar library (XSL; Appendix \ref{xsl}) is overlaid in red. The residuals {\tt (data-model)} are illustrated in green. Two vertical gray lines indicate the spectral region where we actually fit the model templates to the spectra of all bins. Panels C, D, and E: Rotation ($V$), velocity dispersion ($\sigma$), and root-mean-squared velocity ($V_{\rm rms}$=$\sqrt{V^2 + \sigma^2}$).    White contours show the intensity, which decreases by 0.5 magnitudes from the center outward. We also performed the same measurements using the PHOENIX high-spectral-resolution synthetic spectral template (Appendix \ref{phoenix}), which are shown in Fig. \ref{jwst_PHOENIX_maps}.}        
\end{figure*}

This work presents a dynamical stellar-based \Mbh\ measurement for NGC 4736 using data observed by the {\it James Webb} Space Telescope (JWST) Near-Infrared Spectrograph (NIRSpec) G235H/F170LP grating. These high-quality integral field unit (IFU) data enable precise 2D stellar kinematics within a $3\arcsec\times3\arcsec$ field of view (FoV), resolving the longstanding debate on its \Mbh. Our result also demonstrates that JWST\ can probe mass ranges previously thought to be beyond its reach.

\section{Observations and results} \label{obs}

\subsection{JWST IFU} 

NGC~4736 was observed with NIRSpec G235H F170LP (PI: Anil C. Seth; PID: 02016) on February 2, 2023. The data consist of eight exposures, each 218.8 seconds long, for a total of 1750.4 seconds on source. The observation is centered on the galaxy's nucleus, utilized a high-spectral-resolution configuration ($R$$\sim$2700), and covered the wavelength range 1.66--3.17 $\mu$m. The NRSIRS2RAPID readout pattern was used with 14 groups per integration, one integration per exposure, and a four-point medium-cycling dither. The data were reduced following the JWST\ calibration pipeline for NIRSpec IFU using the STScI pipeline v1.8.2 with the Calibration Reference Data System context 1063\footnote{\url{https://jwst-crds.stsci.edu}} \citep{Perna23}. The final cube was produced with a pixel scale of $0\farcs1$, which is 1.5 times smaller than the point spread function (PSF) full width at half maximum (FWHM) of $0\farcs15$ at 2.3~$\mu$m \citep[Fig. 7 of][]{D’Eugenio24NaAs}.  The intensity map of the reduced cube is shown in panel A of Fig. \ref{jwst_kine_ppxf}.

We measured the stellar kinematics from the final cube, using part of the CO band heads' stellar absorptions (2.20--2.36~$\mu$m). We first applied the adaptive \textsc{Voronoi}\footnote{v3.1.5: \url{https://pypi.org/project/vorbin/}} binning method \citep{Cappellari03} to spatially combine fluxes of many spaxels into a single bin, reaching the targeted-S/N = 50 per spectral pixel. Second, we performed log-re-binning along the spectral dimension for each binned spectrum with a {\tt velscale = } 50 \kms\ per pixel calculated using Eq. (8) of \citet{Cappellari17} and then applied the \textsc{Penalized PiXel-Fitting} (\textsc{pPXF}) algorithm\footnote{v8.2.1: \url{https://pypi.org/project/ppxf/}} \citep{Cappellari23} to fit the stellar spectral libraries (Appendix \ref{templates}) to these binned spectra; however, see Appendix \ref{psf_lsf} for our derivation and use of the line-spread function (LSF).

In low-ionization nuclear emission-line region (LINER) galaxies such as NGC~4736, warm dust emission is minimal in the near-infrared  \citep[NIR; see Fig. 7 of][]{Ho08}.  Further evidence of this is provided  by recent JWST/Mid-Infrared Instrument (MIRI) spectra, which show that the dust emission tends to be negligible toward shorter wavelengths \citep[i.e., $<$5 $\mu$m;][]{Goold24} in the nuclear regions of two nearby LLAGNs, Sombrero and NGC 1052. We fit the kinematic parameters $V$ (rotation), $\sigma$ (dispersion), $h_3$ (skewness), and $h_4$ (kurtosis) by setting {\tt moments} = 4, and excluded the additive and multiplicative polynomials by setting {\tt degree} = $-$1 and {\tt mdegree} = 0 \citep{Cappellari23}. We accounted for the instrumental broadening of the NIRSpec spectra by convolving the stellar templates (Appendix \ref{templates}) with the instrumentally different dispersion between the data and the templates. The best-fit template superimposed on sections of the central bin's spectrum is shown in panel B of Fig. \ref{jwst_kine_ppxf}. 

The rotation ($V$), velocity dispersion ($\sigma$), and root-mean-squared velocity ($V_{\rm rms}  = \sqrt{V^2 + \sigma^2}$) are displayed in panels C, D, and E of Fig. \ref{jwst_kine_ppxf}, respectively.   The JWST\ nuclear $\sigma$ measurement is in full agreement with that derived from HST Space Telescope Imaging Spectrograph (STIS) data at a scale of $0\farcs1$ \citep{Pagotto19}, indicating the presence of a central velocity dispersion peak. This is the second clear stellar kinematic signature of a central SMBH, after the Spectrograph for INtegral Field Observations in the Near Infrared (SINFONI) observations of NGC~3640 \citep{Thater19}. It is characterized by a strongly resolved central drop toward the galaxy center and a sharp rise within the innermost spaxels in both the $\sigma$ and $V_{\rm rms}$ maps at the scale of the SMBH's sphere of influence (SOI). The nucleus of NGC 4736 has low rotational characteristics, $V/\sigma \lesssim 0.1,$ across the NIRSpec FoV, but this is likely mainly due to the nearly face-on view of the galaxy disk. The detection in HST\ images of a strong, point-like UV source $\approx$2\farcs{5} northwest of the nucleus led to the intriguing suggestion that the source could be a secondary nuclear black hole, leftover from a galactic merger \citep{Maoz05}. However, our stellar kinematic results do not reveal any signatures at that location.  We validated our kinematic measurements by comparing them with those obtained by the \textsc{pPXF} fit to 1.67--1.83~$\mu$m (using [\ion{Mg}{i}] $\lambda$1.711--1.737 $\mu$m) and 2.13--2.30~$\mu$m (using \ion{S}{i} $\lambda$2.188, 2.183, 2.179 $\mu$m). Our experiments demonstrate that the differences in kinematic measurements among these three fitting approaches are $\lesssim$3\%.

\subsection{HST imaging} 

We used the WFPC2 F814W (pixel scale of 0\farcs1) observations of NGC~4736 taken on July 7, 2002 (Project ID 9042; PI: Smartt) to constrain the galaxy's stellar mass model. These data include two exposures totaling 460 seconds. We also tested the systematics arising from the use of multiband HST\ imaging (see Appendix~\ref{mass_f160w}).

The stellar-light distribution was approximated using a multi-Gaussian expansion (MGE) model\footnote{v5.0: \url{http://purl.com/net/mpfit.}}  \citep{Emsellem94} and the \textsc{Python} version of the {\tt mge\_fit\_sectors\_regularized} procedure \citep{Cappellari02} and assuming a distance of $D = 4.66$ Mpc to NGC~4736 \citep[i.e., the distance to the tip of the red giant branch;][]{Karachentsev03}.   We masked the HST\ image to exclude regions obscured by dust as described in \cite{Nguyen20}. Thus, we decomposed the PSF of the HST/WFPC2 F814W filter, which was computed using the \textsc{TinyTim} package\footnote{\url{https://github.com/spacetelescope/tinytim/releases/tag/7.5}.} \citep{Krist11}, into an \textsc{MGE}. This PSF MGE was then used as an input to the MGE fit of the F814W masked image. This masked HST image was fit with a sum of 2D Gaussians, which were convolved with the PSF MGE, and were then analytically deprojected with a given free inclination ($i$) to obtain a 3D axisymmetric light distribution.  We list the parameters of each spatially deconvolved Gaussian of the \textsc{MGE} model in Table \ref{mgetab} and compare them with the F814W image in the left panel of Fig. \ref{mge_fit_contours}.

\begin{table}[h] 
\centering
\caption{HST/WFPC2 F814W stellar-light MGE model.}  
\footnotesize
\begin{tabular}{cccc}
 \hline\hline
\small $j$ &$\log\Sigma_{\star,j}$ (\Lsun\ ${\rm pc^{-2}})$ &$\log\sigma_j$ ($\arcsec$) &$q'_j=b_j/a_j$\\
\small (1) & (2) & (3) & (4)  \\                        
\hline
\small 1 &  3.867   &$-$0.375&  0.999 \\
\small 2 &  2.811   &$-$0.089&  0.999 \\
\small 3 &  3.894   &  0.198   &  0.950 \\
\small 4 &  3.586   &  0.573   &  0.975 \\
\small 5 &  2.949   &  1.018   &  0.999 \\
\small 6 &  2.345   &  1.325   &  0.999 \\
\small 7 &  2.108   &  1.537   &  0.999 \\
\hline
\end{tabular}\\
\label{mgetab}
\tablefoot{Columns (from left to right): Gaussian component number, surface brightness, dispersion along the major axis, and axial ratios.} 
\end{table}

\begin{figure}[!h]    
    \centering
    \includegraphics[width=0.45\textwidth]{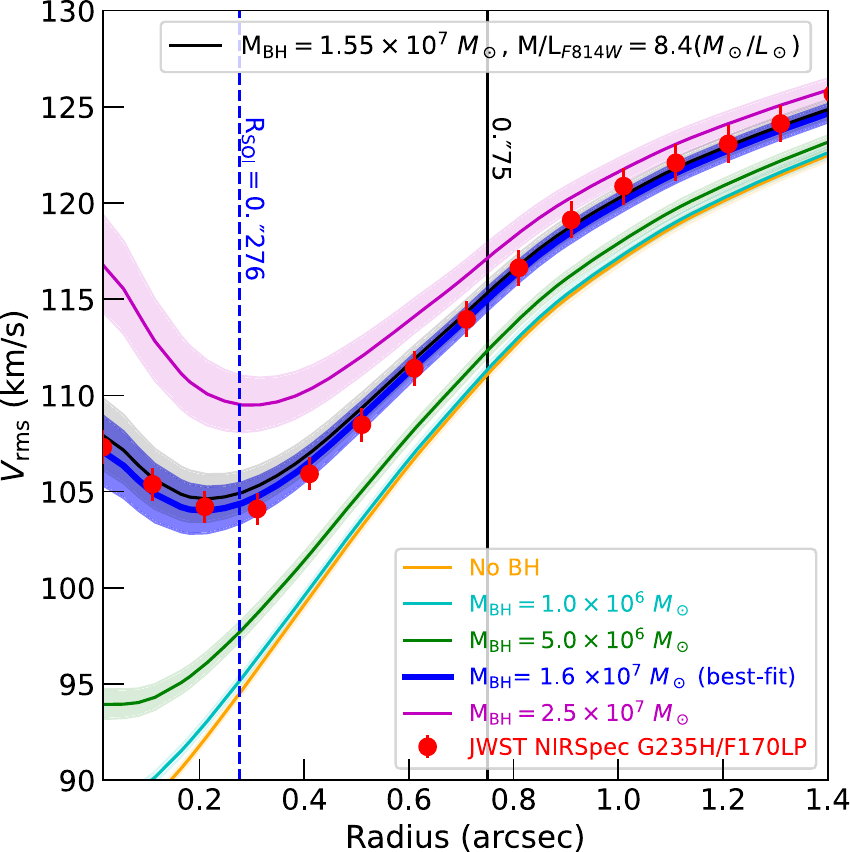} 
     \caption{Comparisons of the stellar kinematics measured from the NIRSpec G235H/F170LP observations and a range of JAMs. The black line shows the model with a constant \ml$_{\rm F814W}$, and colored lines the models that incorporate a radially varying \ml$_{\rm F814W}$ plus 1$\sigma$ error. The dashed vertical line marks the extent of the SMBH's SOI. The data profile was obtained by averaging over circular annuli, and the associated errors are defined as the median of the $V_{\rm rms}$ uncertainties calculated by \textsc{pPXF} (i.e., $\Delta V$ and $\Delta\sigma$) of all bins within the circular annuli.}
    \label{jam_best-fit_1d_corr}
\end{figure}

\section{SMBH mass with the Jeans anisotropic model} \label{bh} 

Using a constant \ml$_{\rm F814W}$ for our stellar-mass model, the initial optimization yielded the best-fit Jeans anisotropic model (JAM; Appendix \ref{jam}) with \Mbh=(1.7$\pm$0.18)$\times10^7$ \Msun\ and \ml$_{\rm F814W}$=8.30$\pm$0.05 \Msun/\Lsun. While this model fits the data well within a radius of $0\farcs75$, discrepancies emerge beyond this radius (left panel of Fig. \ref{jam_best-fit_1d}). We addressed this issue by adjusting \ml$_{\rm F814W}$=8.4 \Msun/\Lsun, improving the agreement between the model and the data from $0\farcs75$ to $2\farcs0$. This correction reduced the \Mbh\ to 1.55$\times$$10^7$~\Msun\ and resolved the discrepancies (see Fig. \ref{jam_best-fit_1d_corr}).

\begin{figure*}[!h]
        \centering
         \includegraphics[width=0.33\textwidth]{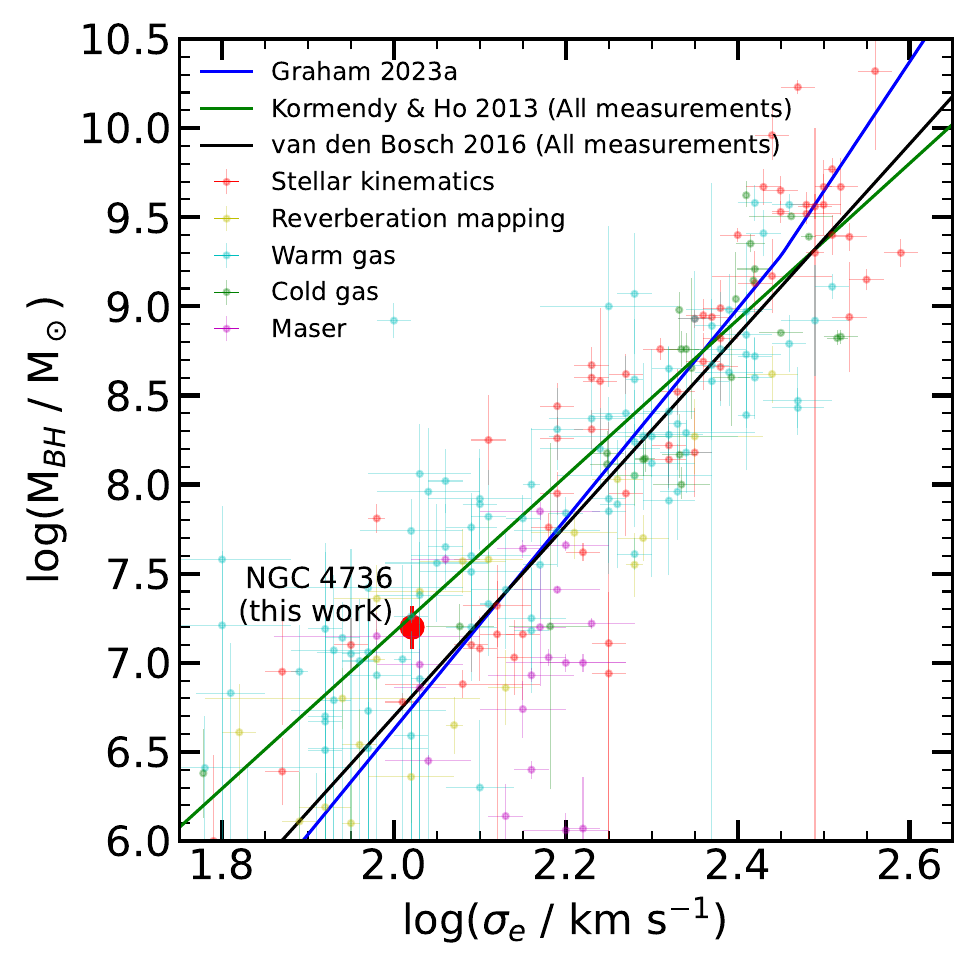} 
         \includegraphics[width=0.33\textwidth]{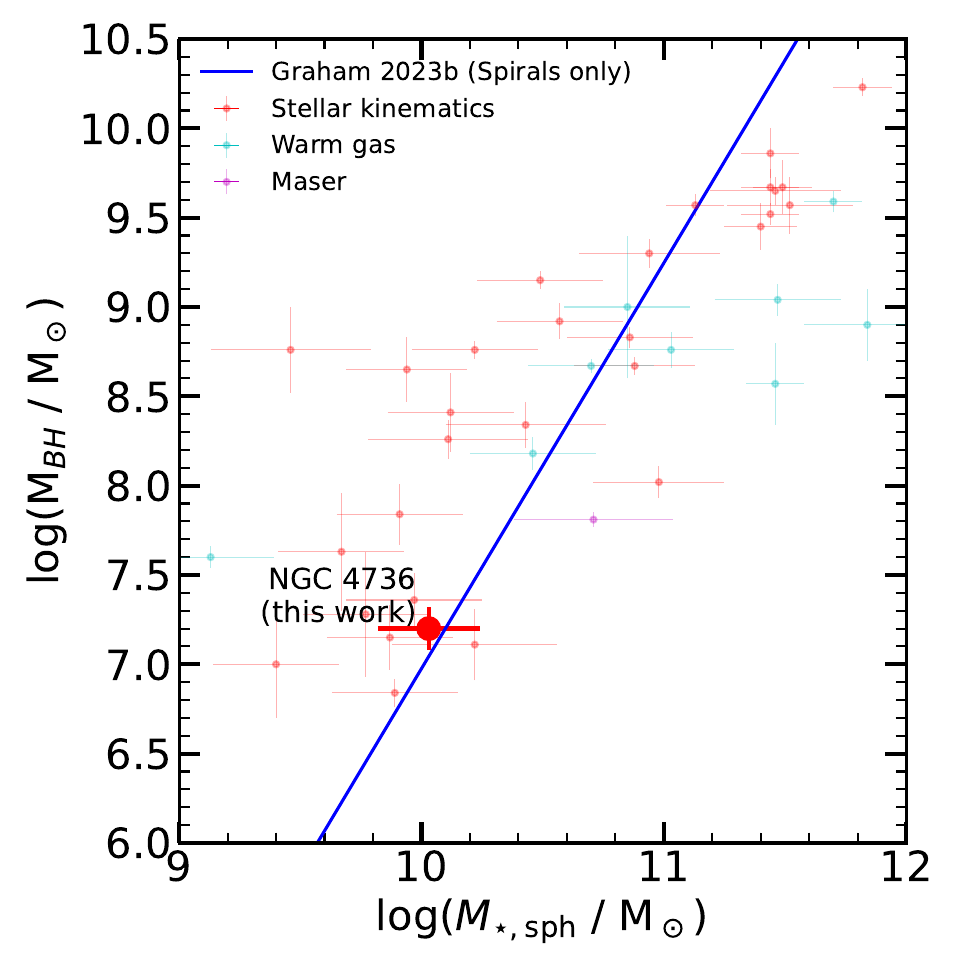} 
         \includegraphics[width=0.33\textwidth]{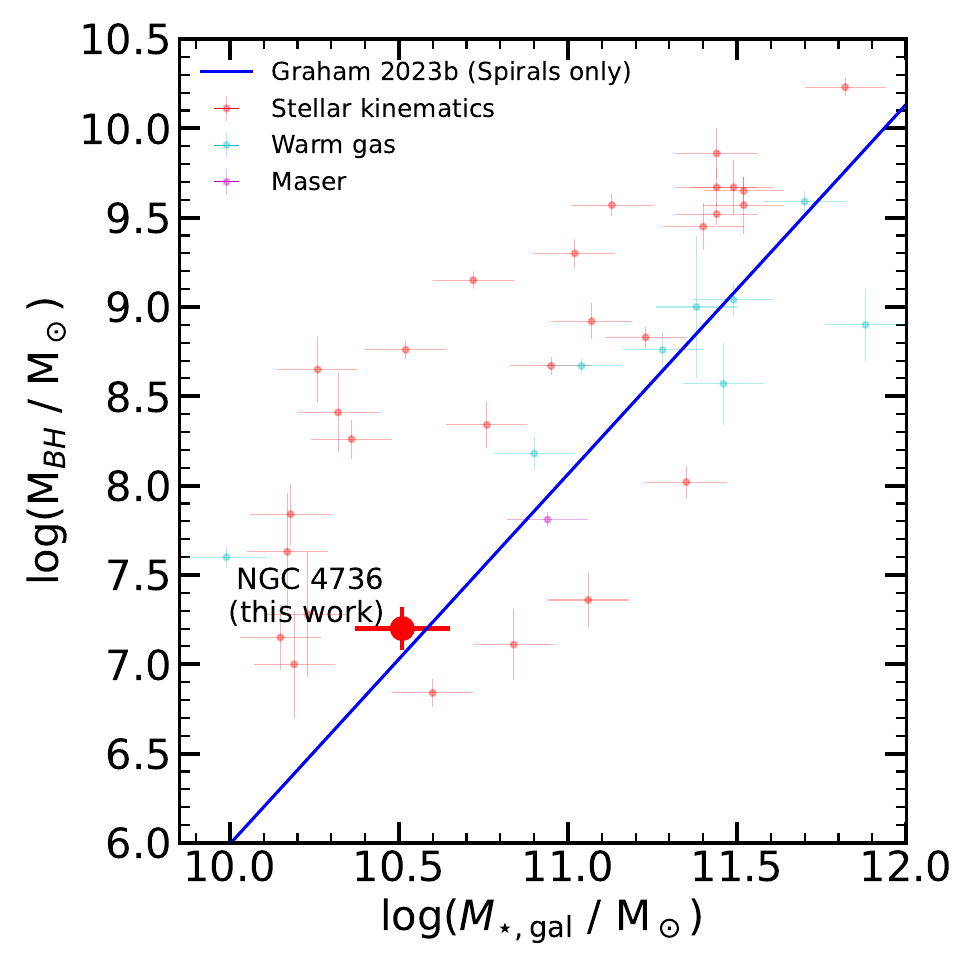}  
        \caption{\label{bhmass_sigma}Location of NGC 4736 with respect to the \Mbh--$\sigma$ (left), \Mbh--$M_{\star, \rm sph}$ (middle), and \Mbh--$M_{\star, \rm gal}$ (right) scaling relations. Our new measurements are consistent within 1$\sigma$ with the \Mbh--$\sigma$ scaling relations for spiral galaxies.}
\end{figure*}

In the second optimization of the stellar mass model, we considered a \ml$_{\rm F814W}$ varying with radius. Specifically, we assumed a  \ml\ profile that is constant at radii $<$0\farcs75 (\ml$_{0\farcs75}$) and $>$2\farcs0 (\ml$_{2\farcs0}$) and varies linearly between the two radii.  The value of (\ml$_{0\farcs75}$) was fixed to 8.3~\Msun/\Lsun\ based on the best-fitting JAM model, while (\ml$_{2\farcs0}$) was left as a free fit parameter.    Thus, the $M/L(\sigma)$ for each Gaussian with dispersion sigma is 
\[
M/L(\sigma) = (M/L)_{0\farcs75} + (\sigma-0\farcs75) \times \left[ \frac{(M/L)_{2\farcs0}-(M/L)_{0\farcs75}}{1\farcs25} \right]. 
\]
We obtained the best-fit constraint with \Mbh=(1.60$\pm$0.16)$\times$$10^7$ \Msun, \ml$_{2\farcs0}$=8.39$\pm$0.04 \Msun/\Lsun, orbital anisotropy $\beta_r$=-0.52$\pm$ 0.03, and $i$=65$^\circ$$\pm$17$^\circ$ (presented in Fig. \ref{jam_best-fit_1d}).  This optimization successfully fits the data across the NIRSpec IFU FoV (Fig. \ref{jam_best-fit_1d_corr}). Table \ref{jamresults} lists the best-fit parameters and associated errors.

Our adopting of a \ml$_{\rm F814W}$ gradient for $0\farcs75 < r < 2\farcs0$ is consistent with a color (F450W–F814W) change in the central region, which is likely caused by a change in the stellar population in the innermost regions, as is clearly visible in multicolor images of this galaxy (Fig. \ref{NGC4736_colormap}).  This adoption of the \ml$_{\rm F814W}$ gradient stems from the fact that we did not calibrate the HST\ images for reddening in this work. This results in an improved fit to the variation in $V_{\rm rms}$ versus radius, and a tighter constraint on \Mbh\ (see Figs. \ref{1D_kinematics_profile_F160W} and \ref{jam_1d_extend} for models that extend beyond the NIRSpec FoV). However, this constrained \ml$_{\rm F814W}$ is 3.5 and 13 times higher than what is predicted by the Salpeter initial mass function using PEGASE-2 models with solar metallicity and stellar ages of 13 Gyr (old) or 3 Gyr (intermediate), respectively; this resulted from the limited kinematic data and the absence of dust correction in our stellar mass model.

Figure \ref{jam_mcmc} shows the 2D scatter plots for each parameter, with colored points representing their likelihood. While the other three parameters are well constrained by JAM, the inclination ($i$) is poorly constrained by the data ($i > 47^\circ$) and is much higher than what was proposed from a morphological study \citep[$i = 35^\circ$;][]{Erwin04}. This could be explained by the slow rotation of the nucleus in NGC~4736, whose nearly spherical kinematics result in a similar appearance across a wide range of inclinations. Notably, the anisotropic orbital parameter constraints indicate that the nucleus of NGC~4736 is strongly dominated by tangential stellar orbits ($\beta_r < 0$).

An unusual feature of our JAM constraints is the positive covariance between \Mbh\ and \ml$_{2\farcs0}$ within the 3$\sigma$ confidence level (CL) region of the 2D probability distribution function (PDF), where \Mbh\ increases with \ml$_{2\farcs0}$. This may be attributable to the continued rise in \ml$_{\rm F814W}$ beyond the NIRSpec FoV, which requires more extended kinematic data for accurate probing. In addition,  the 2D PDF also reveals negative covariances between \Mbh\ and \ml$_{2\farcs0}$ with $\beta_r$ at the same CL, indicating that the distribution of tangential stellar orbits becomes more prominent as \Mbh\ and \ml$_{2\farcs0}$ increase.

\begin{table}
    \centering
     \caption{Best-fit JAM model and uncertainties (HST/F814W).}
     \footnotesize
    \begin{tabular}{|c|c|cc|}
    \hline\hline
 \small (1) & (2) & (3)& (4)   \\ 
\small {\bf Optimization}& Parameters $\equiv$ Initial Parameters & best-fit & 1$\sigma$   \\  
 \hline
\small   {\bf 1}          & $\log_{10}(M_{\rm BH}/$\Msun) $=7$&  7.23  & $\pm{0.04}$   \\
\small(Constant       &\ml$_{\rm F814W}$ (\Msun/\Lsun) = 8  &  8.30  &$\pm{0.05}$ \\
\small\ml)                 &                  $i (^{\circ})$ = 65                  &  65 &$\pm{17}$ \\
 \small                      &                  $\beta_r=-0.5$                     &$-$0.51& $\pm{0.03}$  \\  
\hline   
\small    {\bf 2}       &$\log_{10}(M_{\rm BH}/$\Msun) $=7.23$& 7.21    & $\pm{0.04}$  \\
\small(Varying       & \ml$_{2\farcs0}$ (\Msun/\Lsun) = 8.3    & 8.39   & $\pm{0.04}$  \\
\small\ml)              &                  $i (^{\circ})$ = 65                    &    65   &  $\pm{17}$   \\
\small                    &                  $\beta_r=-0.51$                     &$-$0.52& $\pm{0.03}$  \\ 
\hline    
\end{tabular}
\label{jamresults}
\tablefoot{The search ranges for the model parameters were kept fixed as follows: $0 \leq \log_{10}(\Mbh/\Msun) \leq 9$, $-1 \leq \beta_r \leq 1$, $30^\circ \leq i \leq 90^\circ$, and $0 \leq \ml_{\rm F814W}$ or \ml$_{2\farcs0} \leq 20$~\Msun/\Lsun. Columns 3 and 4 present the best-fit parameters and 1$\sigma$ (16–84\%) errors, respectively.} 
\end{table}

\section{Discussion and summary} \label{discussion} 

Given \Mbh=$1.6 \times 10^7$~\Msun\ \citep{Gultekin19}, a measured central velocity dispersion $\sigma_c$=$105$~\kms\ \citep{Barth02}, and a distance $D$=4.66$\pm$0.59 Mpc \citep{Karachentsev03}, the SMBH’s SOI radius is calculated as $R_{\rm SOI}$=$G\Mbh/\sigma_c^2$=6.2 pc ($0\farcs276$).  This radius is two times larger than the FWHM of NIRSpec's PSF, confirming that our stellar kinematic measurements are well resolved spatially.  The kinematic signature of the SMBH is detected in the innermost $\approx$12 spaxels.    Our  \Mbh\ measurement is the first in external galaxies\footnote{The mass of the SMBH in the Milky Way has also been dynamically measured from the full orbits of individual stars via optical observations \citep{Ghez98}, as well as from radio interferometric modeling of its photon ring \citep{EventHorizonTelescopeCollaborationMW2019}.} accurately estimated using JWST\ observations and the stellar dynamical method.

We find that the model with \Mbh\ = $2.5 \times 10^7$~\Msun\ deviates from the JWST\ data by more than 1$\sigma$. We thus argue that \Mbh\ = $1.6 \times 10^7$~\Msun\ represents the most robust and reliable estimate. In particular, we have presented a model without a black hole along with models that assume \Mbh\ = $10^6$ and $5 \times 10^6$~\Msun\ (see Fig. \ref{jam_best-fit_1d_corr}). The model without a black hole and the one with \Mbh\ = $10^6$~\Msun\ are nearly identical at the center and indistinguishable within the SOI. The model with \Mbh\ = $5 \times 10^6$~\Msun\ begins to rise above the lower-mass models, producing a distinct velocity signature of at least 5~\kms\ at the center, which is more than three times larger than the measurement uncertainties ($\Delta V_{\rm rms}$). This difference is sufficient for the black hole's presence to be detected and measured reliably.

We therefore conservatively conclude that \Mbh$_{\rm , min}$ = $5 \times 10^6$~\Msun\ represents the lowest mass limit accessible with NIRSpec high-spectral-resolution observations using dynamical modeling. However, we note that this limit applies specifically to galaxies with central velocity dispersions $\sigma_c \lesssim 100$~\kms\ and distances $D \lesssim 5$~Mpc.

Our \Mbh\ constraint for NGC~4736 aligns well (within 1$\sigma$) with the \Mbh--$\sigma$ \citep[Fig. \ref{bhmass_sigma};][]{Kormendy13, vandenBosch16, Graham23}, \Mbh--$M_{\star, \rm sph}$ \citep{Graham_Sahu23},  and \Mbh--$M_{\star, \rm gal}$  \citep{Graham24} correlations. While the relations from \citet{Kormendy13} and \citet{vandenBosch16} were derived by grouping spiral, lenticular, and elliptical galaxies, \citet{Graham23a, Graham23} separated the galaxy types, and we used the spiral-galaxy-specific relation. We adopted a galaxy stellar mass of $\log(M_{\star, \rm gal}/$\Msun) $= 10.51\pm0.14$ and a spherical-bulge stellar mass of $\log(M_{\star, \rm sph}/$\Msun) $= 10.03\pm0.21$ \citep{Graham_Sahu23} based on the galaxy-light decomposition and a diet-Salpeter initial mass function \citep{DavisBL19}. 

Our inferred \Mbh\ corresponds to a bolometric-to-Eddington luminosity ratio of $L_{\rm bol}/L_{\rm Edd} \approx 1.1 \times 10^{-5}$, which we determined using the value of $L_{\rm bol}=2.2 \times 10^{40}$ erg s$^{-1}$ derived by \citet{Constantin12} and rescaled to our adopted distance. This ratio places our  \Mbh\ within the range typical for LINER targets \citep{Ho08}. Our \Mbh\ is consistent with the value derived from the updated FP of black hole activity \citep{Gultekin19}, which is applicable to LLAGNs. This agreement resolves the longstanding debate on the \Mbh\ estimates for NGC~4736, as our result aligns with the predictions from the $M_{\rm BH}$–$\sigma$ and $M_{\rm BH}$–$M_\star$ relations for spiral galaxies.

 \begin{acknowledgements}
 
We thank the referee for useful comments and discussion that helped improve the quality of the manuscript. 
We would like to thank the ICISE staff for their enthusiastic support.
This work is partially supported by a grant from the Simons Foundation to IFIRSE, ICISE (916424, N.H.). 
This research was funded by Vingroup Innovation Foundation (VINIF) under project code VINIF.2023.DA.057.
RS acknowledges support from the INAF grant number 1.05.23.04.04.
NT would like to acknowledge partial support from UKRI grant ST/X002322/1 for UK ELT Instrument Development at Oxford. IC’s research is supported by the SAO Telescope Data Center. He also acknowledges support from the NASA ADAP-22-0102 and HST-GO-16739 grants. MPS acknowledges support under grants RYC2021-033094-I, CNS2023-145506 and PID2023-146667NB-I00 funded by MCIN/AEI/10.13039/501100011033 and the European Union NextGenerationEU/PRTR. AWG thanks Bloomfield Higher Secondary School, Zunheboto, Nagaland, India, from where he worked. 

\end{acknowledgements}

\bibliography{aa54672-25}
\bibliographystyle{aa}

\begin{appendix}

\section{Stellar spectral libraries}\label{templates}

\begin{figure*}[!h]
        \centering
        \includegraphics[width=0.65\textwidth]{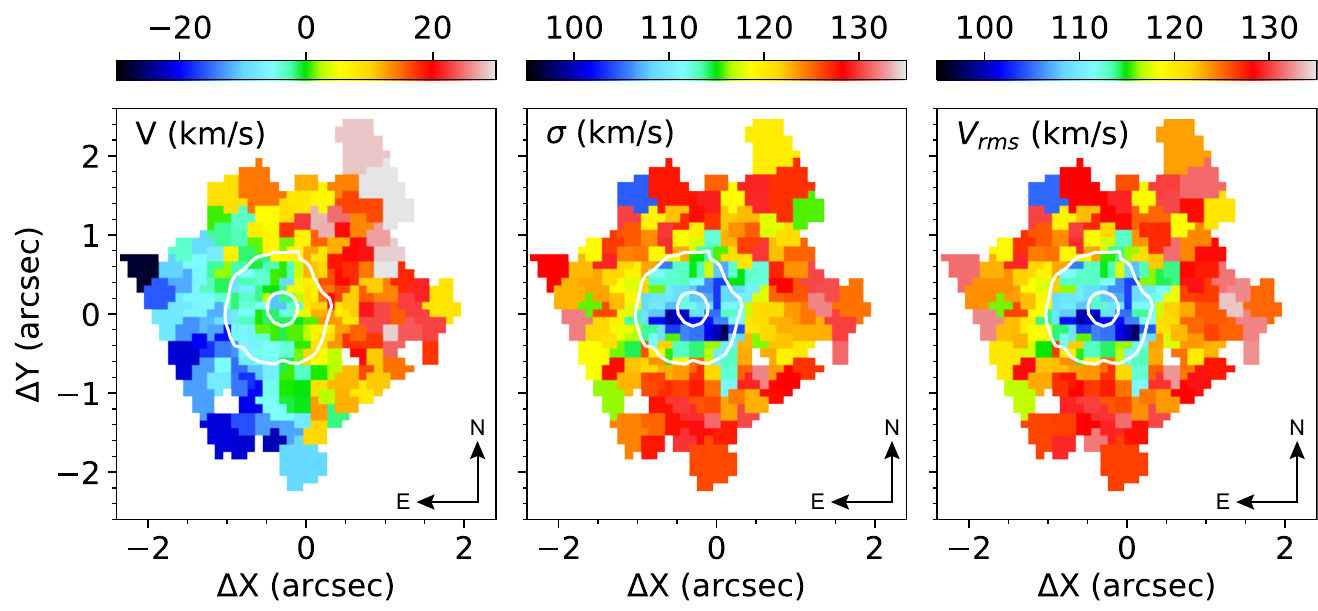} \hspace{5mm}
        \parbox[b]{0.31\textwidth}{
        \caption{Same as panels C, D, and E in Fig. \ref{jwst_kine_ppxf}, but showing kinematics derived with the PHOENIX high-spectral-resolution synthetic spectral template, providing a comparison to the results obtained with the XSL. 
         \label{jwst_PHOENIX_maps}}}
\end{figure*}

Stellar spectral libraries are essential tools for analyzing observed spectra and have broad applications in astrophysics. In this work, we employed two distinct types of stellar spectral libraries — one empirical, i.e., the X-shooter stellar library (XSL) and the other synthetic, i.e., the Extensive library of PHOENIX stellar atmospheres and synthetic spectra — to extract stellar kinematics from NIRSpec G235H/F170LP observations. As each type of library has its respective strengths and limitations, it is crucial to evaluate the robustness of our stellar kinematics extraction by comparing results derived from both.

Empirical libraries provide spectra based on real observations, ensuring accurate representation of physical phenomena. For instance, absorption features due to atomic and molecular transitions are reproduced at the correct wavelengths with appropriate shapes and strengths. However, these libraries are inherently limited in spectral resolution and wavelength coverage. Extending empirical libraries often requires significant observational efforts and is constrained by instrumental capabilities. Moreover, the chemical compositions of the stars within such libraries are not precisely known, potentially introducing systematic uncertainties in abundance determinations. Additionally, the parameter space spanned by empirical libraries is often restricted, typically covering a limited range of effective temperatures, surface gravities, and metallicities.

In contrast, synthetic spectral libraries are constrained by the completeness of atomic and molecular line lists, uncertainties in line-broadening parameters, and assumptions in modeling, such as plane-parallel versus spherical geometries or local thermodynamic equilibrium (LTE) versus non-LTE conditions. Despite these limitations, synthetic libraries offer significant advantages: they can be tailored to span a wide range of stellar parameters, elemental abundances, wavelength regimes, and spectral resolutions. These flexible capabilities make synthetic libraries valuable for a variety of astrophysical analyses. 

Our stellar kinematic derivations in the nucleus of NGC 4736, using two stellar spectral libraries (XSL and PHOENIX) with the NIRSpec G235H/F170LP IFU and \textsc{pPXF} (as detailed in Sect. \ref{obs}), show consistency between the results within 5\% when comparing the Light-of-Sight Velocity Distribution (LOSVD) maps in Fig. \ref{jwst_kine_ppxf} (XSL) to those in Fig. \ref{jwst_PHOENIX_maps} (PHOENIX).  However, it should be noted that for the subsequent analysis (i.e., SMBH mass estimates), we adopted the stellar kinematics derived using the XSL library as the canonical reference measurement, while those measured using the PHOENIX library were used for cross-checking.

We have demonstrated that our \textsc{pPXF} stellar kinematic measurements from the NIRSpec G235H/F170LP using the empirical stellar population template XSL, are consistent with those derived using the synthetic stellar population model PHOENIX. Previous studies have also reported similar consistency when deriving the nuclear stellar kinematics of the ultracompact dwarf galaxy M60-UCD1 \citep{Seth14} and a sample of nearby dwarf elliptical galaxies \citep{Nguyen18}, based on Gemini/NIFS and VLT/SINFONI data, using the synthetic PHOENIX templates and the empirical \citet{Wallace96} library of high-resolution spectra of ordinary cool stars.

\subsection{X-shooter stellar library (XSL)}\label{xsl} 

We used the third data release (DR3\footnote{\url{http://xsl.u-strasbg.fr}}) of the X-shooter stellar library \citep{Verro22}, which includes 830 stellar spectral templates from 683 stars. The DR3 spectra covers the full wavelength range of the X-shooter spectrograph, spanning 3000--25,000 \AA, with a resolution of $R\approx10,000$ and dust extinction corrected. The library includes a broad range of stellar types, from O to M, including asymptotic giant branch stars, and covers most of the Hertzsprung--Russell diagram.

\subsection{Library of PHOENIX stellar synthetic spectra}\label{phoenix} 

We cross-checked our stellar kinematics measurements obtained with \textsc{pPXF} using an alternative high-resolution synthetic spectral library, PHOENIX\footnote{\url{http://phoenix.astro.physik.uni-goettingen.de}} \citep{Husser_2013} of 11 stellar templates. Although the templates cover a spectral range of 500--55,000 \AA, when compared to XSL in the NIR, PHOENIX offers 50 times higher spectral resolution ($R\approx500,000$), extends from 0.3 to 5 $\mu$m, and lacks gaps associated with telluric absorption bands. The library covers a parameter space of 2,300 K $\leq T_{\rm eff} \leq 25,000$ K, $0.0 \leq \log(g) \leq +6.0$, $-4.0 \leq [{\rm Fe/H}] \leq +1.0$, and $-0.2 \leq [\alpha/{\rm Fe}] \leq +1.2$. The synthetic stellar spectra were generated from the spherical mode of PHOENIX, which was employed to create model atmospheres, accounting for micro-turbulence in the stellar atmospheres.

\section{Line spread function for NIRSpec}\label{psf_lsf}

In Sect. \ref{obs} we adopt a constant Gaussian LSF with an instrument broadening of $\sigma_{\rm LSF}$ = {\tt velscale =} 50 \kms\ per pixel \citep{Cappellari17}, (or FWHM of 117 \kms\ per pixel at 2.3~$\mu$m) to derive stellar kinematics via \textsc{pPXF} with the NIRSpec G235H/F170LP IFU ($R=2700$). Here, we double check this assumption by re-deriving the LSF using the JWST/NIRSpec IFU spectrum of the standard star P330E (PID: 01538, PI: Gordon, Karl D.). P330E is a a solar-type star (G2V) with many absorption lines and has a relatively low velocity dispersion that's necessary to measure the LSF.

The LSF may vary across the FoV of NIRSpec G235H  F170LP, as has been observed in previous studies using Gemini/NIFS \citep{Seth10} and VLT/SINFONI \citep{Nguyen18}. Here, we characterize potential variations in the LSF by fitting a high-resolution synthetic template (i.e., XSL or PHOENIX) to the JWST/NIRSpec spectrum of the calibration star P330E. This is done by convolving the template with a Gaussian kernel and performing a $\chi^2$ minimization to determine the instrumental broadening. Given the context of deriving stellar kinematics using the CO band head region, we particularly fit one or more of the following absorption features: the $^{12}$CO(2–0) band head at 2.2935 µm, which is very strong in cooler stars; the \ion{Na}{i} doublet at 2.2062 and 2.2090 $\mu$m, moderately strong in G–M stars and relatively sharp and isolated; the \ion{Ca}{i} triplet at 2.261, 2.263, and 2.265 $\mu$m, commonly found in cool dwarfs and giants; and the \ion{Mg}{i} line at 2.2814 $\mu$m, which can be blended but is sometimes strong depending on metallicity. Once they reach the spectrograph, these absorption features are dispersed such that their intensity distribution with wavelength reflects the instrument’s LSF. Assuming the central wavelengths of these absorption lines (corrected for small redshifts) are known, we treat their Gaussian profiles as representative of the LSF. After locating these absorptions, we define a region around each absorption, subtract the continuum, normalize the absorption flux to the deep value, and sum the profiles to characterize the LSF.

The spectral resolution across the detector has a median value of 8.52 \AA\ FWHM at 2.3 $\mu$m ($R$$=$2700), with values ranging from 7 to 11 \AA\ FWHM across the G235H/F170LP wavelength range of 16,600–31,700 \AA. The final step before extracting the stellar kinematics is to apply the same spatial binning to the LSF cube as was used for the final IFU of NGC~4736. This varying LSF is then incorporated into the kinematic extraction using \textsc{pPXF}.  We found that using the spatially varying LSF, instead of the constant broadening (Sect. \ref{obs}), did not significantly affect the results, as the velocity dispersion is relatively high.

\section{Change in populations in the mass model}\label{pop_changes}

\begin{figure*}[!h]    
    \centering
    \includegraphics[width=0.45\textwidth]{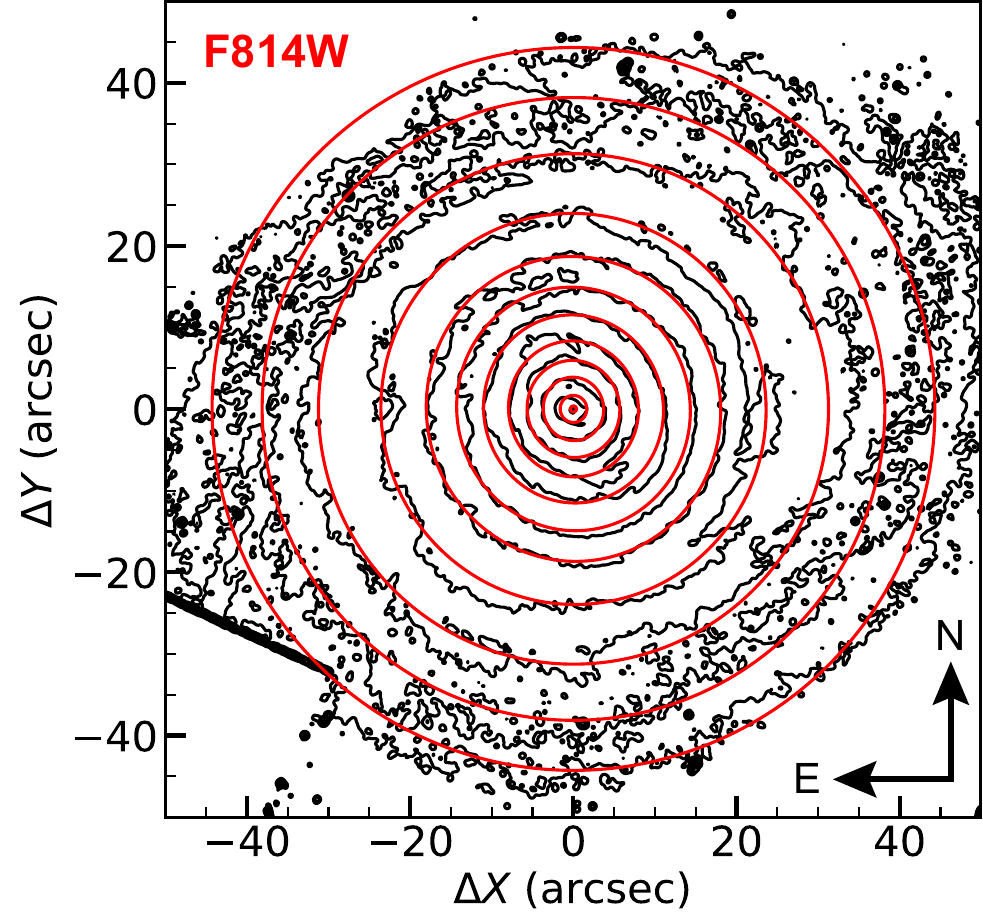}
    \hspace{5mm}
    \includegraphics[width=0.45\textwidth]{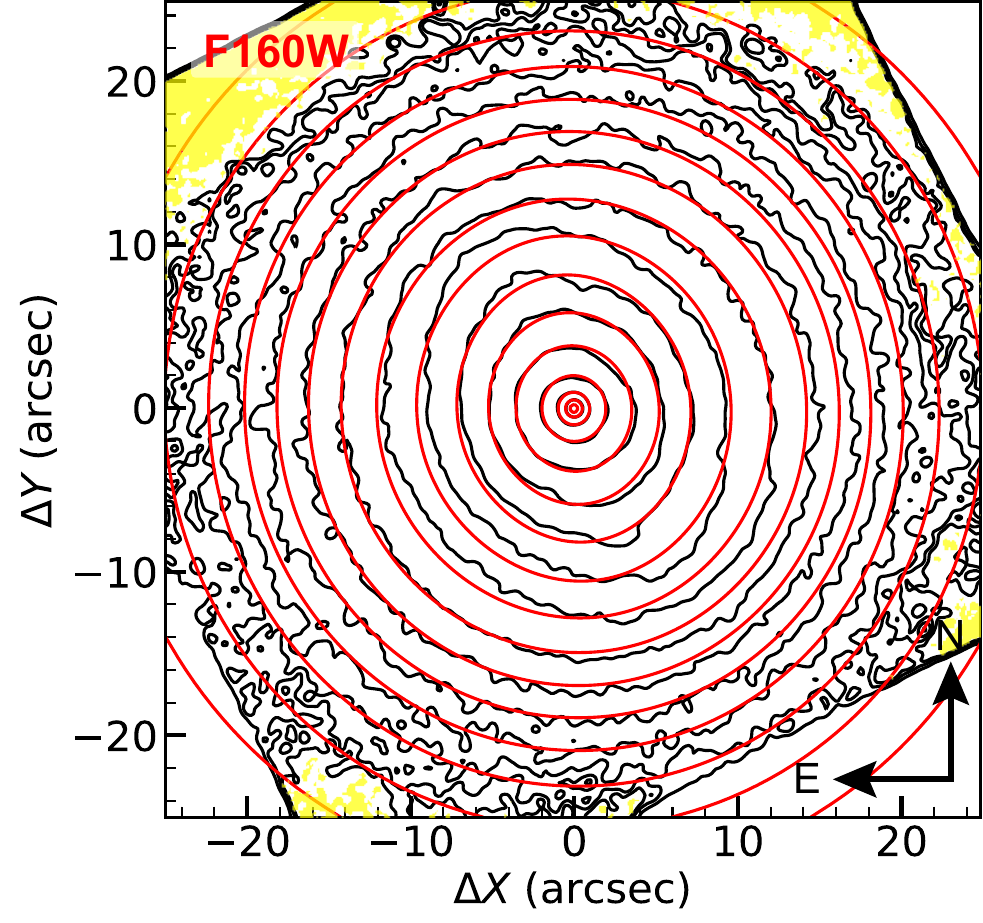}
     \caption{Comparison of the HST/WFPC2 F814W ({left}) and HST/NICMOS NIC3 F160W images ({right}) with their MGE models, shown in 2D surface brightness density. Black contours represent the data and red contours the model, highlighting their alignment at corresponding radii and contour levels.}
    \label{mge_fit_contours} 
\end{figure*}

\begin{figure*}[!h]
    \centering
        \includegraphics[width=0.9\textwidth]{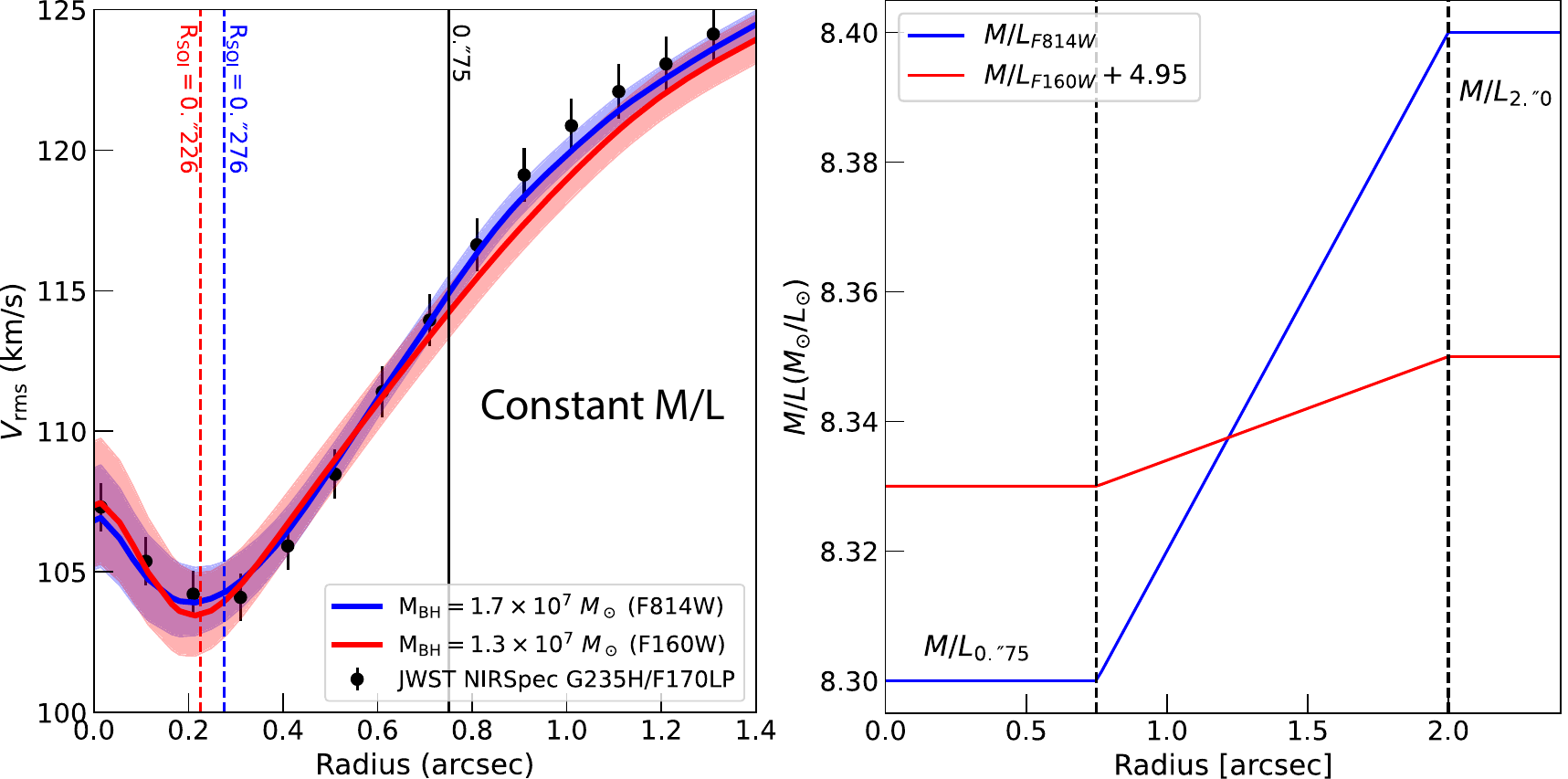} 
        \caption{\label{jam_best-fit_1d}{Left:} Comparison of the stellar kinematics measured from NIRSpec G235H/F170LP observations and the best-fit JAM (with 1$\sigma$ error) with a constant \ml$_{\rm F814W}$ (solid blue line) and \ml$_{\rm F160W}$ (solid red line). Discrepancies between the data and model are apparent beyond $0\farcs75$, as marked by the dash vertical line.   {Right:} Best-fit radially varying \ml$_{\rm F814W}$ profile (described in  Sect. \ref{obs}; solid blue line).\ It exhibits a step-like characteristic with $M/L_{0\farcs75} = 8.3$~\Msun/\Lsun\ at $r \leq 0\farcs75$ and $M/L_{2\farcs0} = 8.39~$\Msun/\Lsun\ at $r \geq 2\farcs0$. Between these radii, the \ml$_{\rm F814W}$ profile increases linearly. Similarly, for the best-fit radially varying \ml$_{\rm F160W}$ profile (Appendix~\ref{mass_f160w}; solid red line), we adopt $M/L_{0\farcs75} = 3.38~$\Msun/\Lsun\ at $r \leq 0\farcs75$ and $M/L_{2\farcs0} = 3.40~$\Msun/\Lsun\ at $r \geq 2\farcs0$. Here, the \ml$_{\rm F160W}$ profile is vertically offset by a value of 4.95~\Msun/\Lsun\ for visualization.}    
\end{figure*}

\begin{figure}[!h]    
    \centering
    \includegraphics[width=0.45\textwidth]{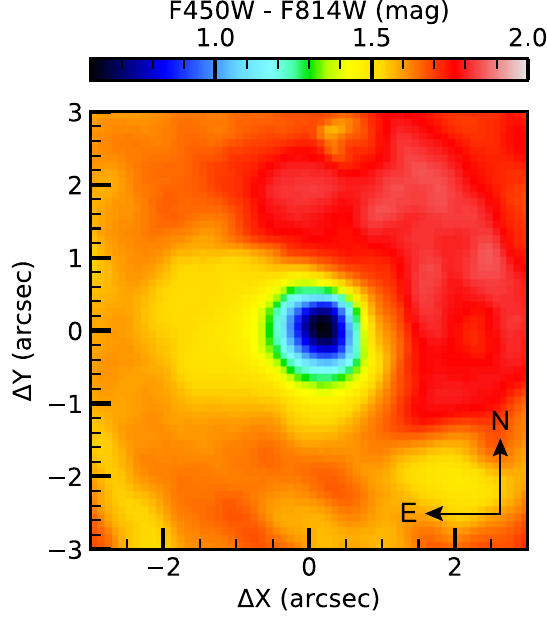}
     \caption{Central $6\arcsec \times 6\arcsec$ color map of NGC~4736, derived from HST/WFPC2 F450W and F814W images.\ We see a significant variation in the range $0\farcs75 < r < 2\farcs0$. This gradient is likely attributed to a transition in stellar populations within this region, supporting our adoption a radially varying \ml$_{\rm F814W}$ profile in the galaxy mass model.}
    \label{NGC4736_colormap}
\end{figure}

Our dynamical model identified evidence for a subtle variation in the \ml$_{\rm F814W}$ profile within the annular region of 0\farcs75$<$$r$$<$2\farcs0, where the profile exhibits a linear increase (right panel of Fig. \ref{jam_best-fit_1d}). To investigate the origin of this variation, we examined the corresponding changes in the HST/WFPC2 (F450W–F814W) color. We constructed central $6\arcsec \times 6\arcsec$ color map of NGC~4736, shown in Fig. \ref{NGC4736_colormap}, by converting the fluxes from the F450W and F814W images (measured in counts s$^{-1}$) into their respective 2D surface brightness maps (in units of mag arcsec$^{-2}$).

To ensure the accuracy of the color map, we applied a cross-convolution technique, where each image was convolved with the PSF of the other filter (e.g., for the F450W–F814W color map, the F450W image was convolved with the F814W PSF, and vice versa). This approach mitigates artificial gradients near the galaxy’s center caused by differences in PSF widths between the filters. Subsequently, the background level for each image was determined in an annulus located at the maximum radius available in the observations (beyond $40\arcsec$ from the nucleus) and subtracted from the images. Finally, the (F450W–F814W) color map was created using the ABmag-based zero points, with corrections applied for foreground extinction.

Variations in the color map clearly reveal signatures of two distinct and dominant stellar populations: younger stars in the nuclear region and older stars at larger radii. A transition in color is observed within the annular region of $0\farcs75 < r < 2\farcs0$, consistent with our finding of a linear increase in the \ml$_{\rm F814W}$ profile.

Furthermore, Fig. \ref{NGC4736_colormap} shows that the color gradients extend clearly into the very central pixels (i.e., $r \lesssim 0\farcs75$), indicating the presence of kinematically colder stellar populations toward the nucleus. This suggests that the appropriate radial $M/L$ gradient should likely extend all the way to the center as well, which would, in turn, lead to a higher inferred \Mbh. A more detailed investigation of this effect, which would require significantly deeper and more expensive observations, is beyond the scope of this paper and is planned for future work, using stellar population analysis based on Keck optical spectroscopy obtained with OSIRIS at a spatial scale of $\approx$$0\farcs035$ for the nucleus of NGC~4736.

\section{Mass model with the HST/F160W image}\label{mass_f160w}

\begin{figure*}[!h]
\centering
    \includegraphics[width=0.45\textwidth]{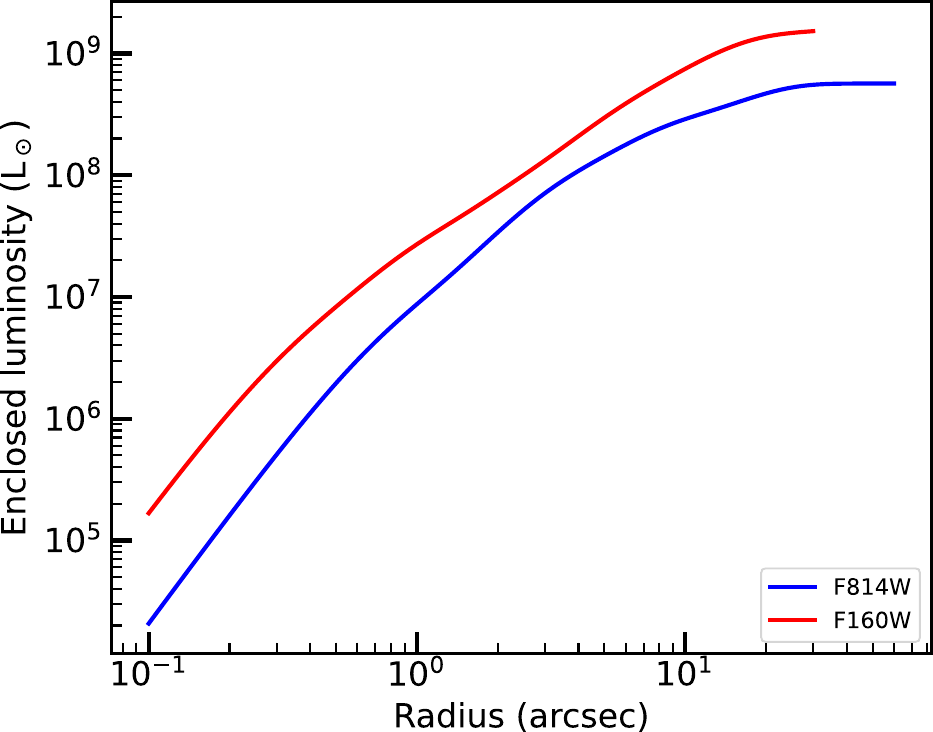}\hspace{2mm}
    \includegraphics[width=0.45\textwidth]{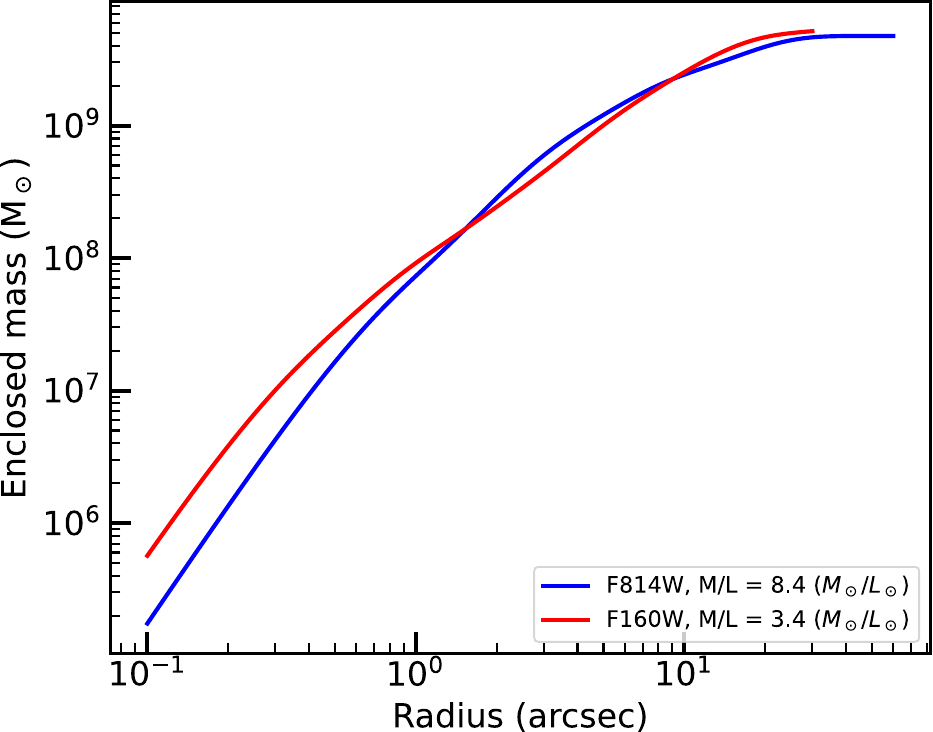}\
        \caption{\label{N4736_enclosed_mass_compare}Enclosed luminosity ({left}) and mass ({right}) profiles constructed from the stellar-light MGE models derived from the HST/WFPC2 F814W image, assuming \ml$_{\rm F814W} = 8.4$ \Msun/\Lsun\ (Table~\ref{mgetab}), and from the HST/NICMOS NIC3 F160W image, assuming \ml$_{\rm F160W} = 3.4$ \Msun/\Lsun\ (Table~\ref{mgetab_F160W}). Profiles are plotted at their widest radial extends, i.e., $\approx$50\arcsec\ for the WFPC2 F814W and $\approx$25\arcsec\ for the NICMOS F160W images.}
\end{figure*}

As we did not apply a reddening correction to the HST/WFPC2 F814W image used to derive the stellar-mass model in Sect. \ref{obs},  this effect is likely accounted for the radially varying \ml$_{\rm F814W}$  profile defined in Sect. \ref{bh}. We addressed this systematic effect by using HST\ imaging in multiple bands. In particular, we cross-checked our stellar-mass model using the HST/NICMOS NIC3 F160W broad-band image at 1.6~$\mu$m, which also has a pixel scale of 0\farcs1. The data were taken on 2003 April 20 (Project ID 9360; PI: Kennicutt). They consist of four exposures of 96 seconds each, totaling 384 seconds.

We derived the stellar-light distribution of NGC~4736 from this NIR imaging data using the MGE technique, following the same procedure described in Sect. \ref{obs}. In this process, we also modeled the PSF of the NICMOS NIC3 F160W filter using \textsc{TinyTim} \citep{Krist11}, then decomposed it with the image for the MGE. The parameters of each spatially deconvolved Gaussian component of the \textsc{MGE} model are listed in Table~\ref{mgetab_F160W}, and the comparison with its F160W image is shown in the right panel of Fig.~\ref{mge_fit_contours}. It is clear that there is less reddening in the F160W image than in the F814W image, as evidenced by the better agreement between the data and the MGE model across the FoV of the F160W image.

Figure~\ref{N4736_enclosed_mass_compare} compares the cumulative luminosity, calculated from Table~\ref{mgetab} for F814W and Table~\ref{mgetab_F160W} for F160W, and the corresponding stellar mass, derived from these luminosities using the constant \ml\ values \ml$_{\rm F814W} = 8.4$ \Msun/\Lsun\ (from Sect.~\ref{bh}) and \ml$_{\rm F160W} = 3.4$ \Msun/\Lsun\ (from Appendix~\ref{jam}). The increased light in the NIR filter is compensated by the lower \ml\ returned from the dynamical models. However, there is more central mass within 2\arcsec\ in the F160W mass model compared to that derived from the F814W mass model, resulting in a smaller \Mbh.

\begin{table}[!h] 
\centering
\caption{HST/NICMOS NIC3 F160W stellar-light MGE model.}  
\footnotesize
\begin{tabular}{cccc}
 \hline\hline
\small $j$ &$\log\Sigma_{\star,j}$ (\Lsun\ ${\rm pc^{-2}})$ &$\log\sigma_j$ ($\arcsec$) &$q'_j=b_j/a_j$\\
\small (1) & (2) & (3) & (4)  \\                        
\hline
\small 1 &  4.386   &$-$0.742&  0.999 \\
\small 2 &  4.434   &$-$0.339&  0.950 \\
\small 3 &  4.002   &  0.006   &  0.999 \\
\small 4 &  3.914   &  0.409   &  0.999 \\
\small 5 &  3.563   &  0.664   &  0.700 \\
\small 6 &  3.726   &  0.891   &  0.955 \\
\small 7 &  2.375   &  1.240   &  0.999 \\
\hline
\end{tabular}\\
\label{mgetab_F160W}
\parbox[t]{0.49\textwidth}{\small \textit{Notes:} Columns in order: Gaussian component number, surface brightness, dispersion along the major axis, and axial ratios.} 
\end{table}

\begin{figure}[!ht]    
    \centering
    \includegraphics[width=0.45\textwidth]{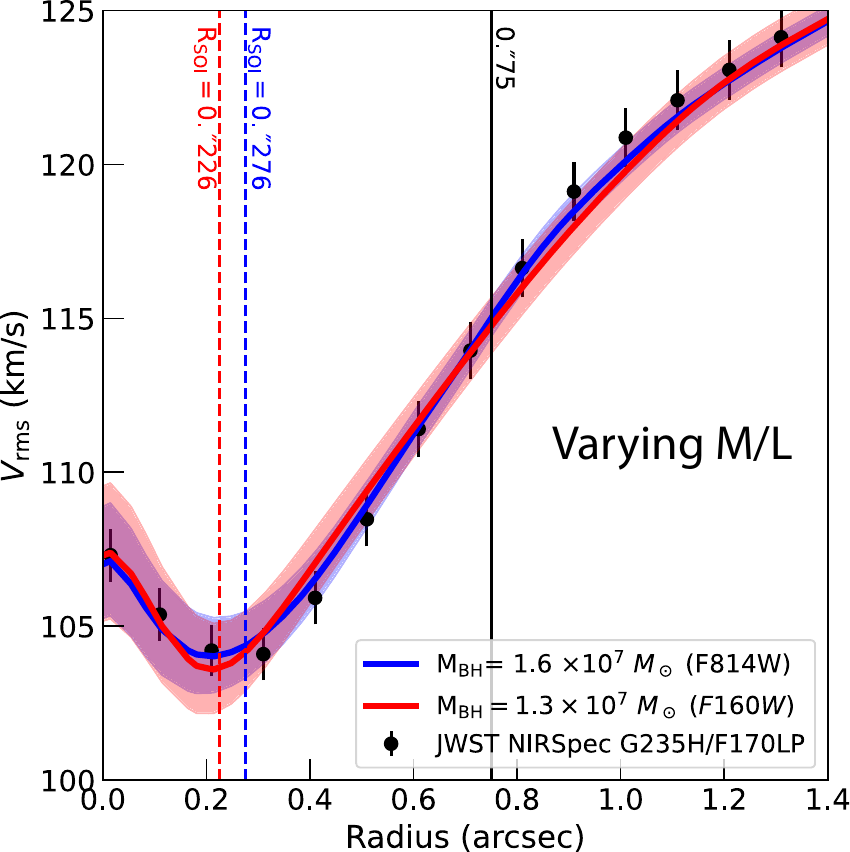} 
     \caption{Comparison of the 1D $V_{\rm rms}$ profiles from the best-fit JAM model with $M_{\rm BH} = 1.6 \times 10^7$ \Msun\ and the radial \ml$_{\rm F814W}$ variation profile (solid blue line), and an alternative model with $M_{\rm BH} = 1.3 \times 10^7$ \Msun\ and the radial \ml$_{\rm F160W}$ variation profile (solid red line), within the radial extent of the NIRSpec FoV.} 
    \label{1D_kinematics_profile_F160W}
\end{figure}

As seen in the left panel of Fig.~\ref{jam_best-fit_1d}, the constant \ml\ JAM models based on both stellar-mass models from the WFPC2 F814W and NICMOS F160W images fail to provide good fits beyond 0\farcs75, although the former stellar-mass model shows a noticeable improvement in matching the data. We therefore applied the radially varying \ml$_{\rm F160W}$ profile described in Sect.~\ref{bh}, with $M/L(\sigma) = (M/L)_{0\farcs75} = 3.38~$\Msun/\Lsun\ within $0\farcs75$, where $\sigma$ is taken from Table~\ref{mgetab_F160W}. We show the best-fit \ml$_{\rm F160W}$ variation profile in the right panel of Fig.~\ref{jam_best-fit_1d}, together with the same \ml$_{\rm F814W}$ profile, which is vertically offset by a constant value of 4.95~\Msun/\Lsun\ for visualization purposes. Although the linear increase of the \ml\ profile within the annulus $0\farcs75 < r < 2\farcs0$ is small, it does significantly improve the fit between the model and the data beyond $0\farcs75$, as clearly seen by comparing the left panel of Fig.~\ref{jam_best-fit_1d} with Fig.~\ref{1D_kinematics_profile_F160W}. Moreover, the change in \Mbh\ when switching the mass model from a constant \ml$_{\rm F160W}$ to one with a linear variation profile is negligible, indicating that reddening has a minimal effect on our dynamical modelling at NIR wavelengths.

\begin{figure}[!ht]    
    \centering
    \includegraphics[width=0.45\textwidth]{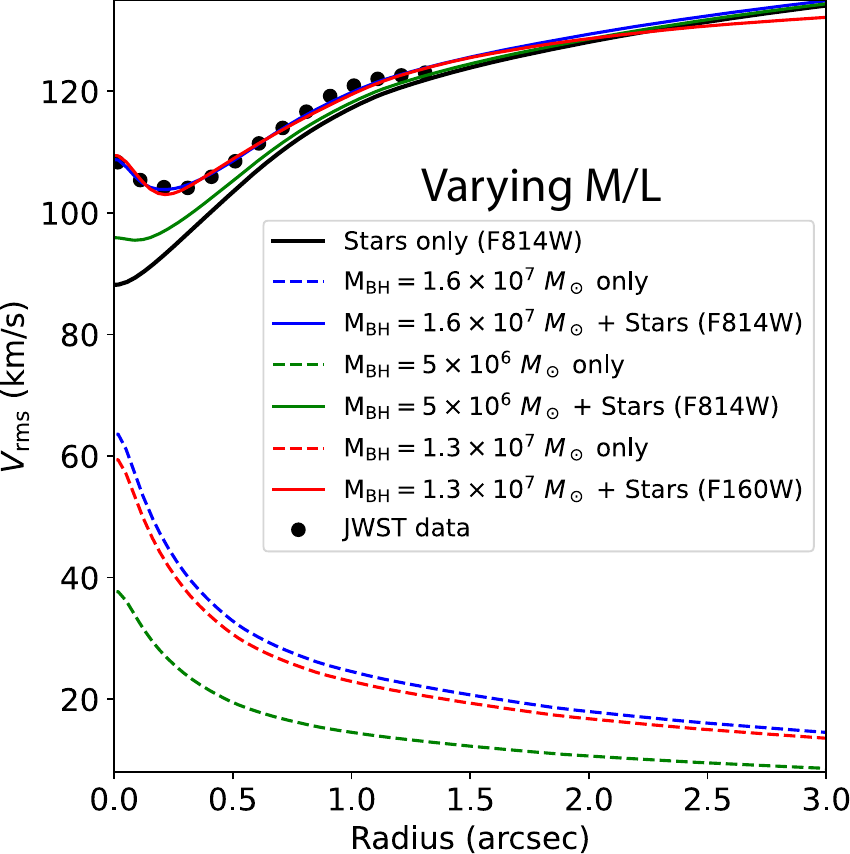} 
     \caption{Comparison of the 1D $V_{\rm rms}$ profiles from the best-fit JAM model with radially varying \ml\ profiles and  $M_{\rm BH} = 1.6 \times 10^7$ \Msun\ (using the F814W stellar-mass model; solid red line, $R_{\rm SOI} \approx0\farcs276$) or $M_{\rm BH} = 1.3 \times 10^7$ \Msun\ (using the F160W stellar-mass model; solid blue line, $R_{\rm SOI} \approx 0\farcs226$) and an alternative model with $M_{\rm BH} = 5 \times 10^6$ \Msun\ (solid green line) against the JAM model with $M_{\rm BH} = 0$ \Msun\ (solid black line) within a radial extent of $3\arcsec$. Despite the small SOI of the SMBH ($R_{\rm SOI} \approx 0\farcs226-0\farcs276$), our best-fit \Mbh\ model exhibits a kinematic imprint extending across the NIRSpec FoV ($\approx$$1\farcs5$) compared to the JAM model without a black hole (which accounts only for the stellar mass component). This results in a systematic increase in $V_{\rm rms}$ by approximately $\Delta V_{\rm rms} \approx 2$ \kms, with the shift becoming more pronounced for higher \Mbh\ values, as shown in Figs. \ref{jam_best-fit_1d_corr}  and \ref{1D_kinematics_profile_F160W}. For reference, an SMBH at the lowest mass limit detectable by NIRSpec high-spectral-resolution observations using stellar dynamical techniques ($M_{\rm BH, min} = 5 \times 10^6$ \Msun\, claimed in this work; Sect. \ref{discussion}) does not produce a visible kinematic effect beyond $\approx$$0\farcs8$.} 
    \label{jam_1d_extend}
\end{figure}

In Fig.~\ref{1D_kinematics_profile_F160W} we show the 1D $V_{\rm rms}$ profiles produced from the best-fit JAM models with varying \ml\ for both the F814W and F160W stellar mass models. Both models provide good fits to the stellar kinematic data across the NIRSpec FoV. However, the \Mbh\ constrained from the F160W mass model is 19\% lower than that derived from the F814W mass model, consistent within the 2$\sigma$ uncertainty.

Figure~\ref{jam_1d_extend} again presents the best-fit models with varying \ml, taken from Fig.~\ref{1D_kinematics_profile_F160W}, to demonstrate that they accurately reproduce the NIRSpec stellar kinematics data. We also show additional models without a black hole or with the minimum detectable black hole mass, \Mbh$_{\rm , min} = 5 \times 10^6$~\Msun\ that could be accessible with NIRSpec high-spectral-resolution observations. This dynamical technique is strictly applicable to galaxies with central velocity dispersions $\sigma_c \lesssim 100$~\kms\ and distances $D \lesssim 5$~Mpc.

\section{Jeans anisotropic model}\label{jam}

\begin{table}
    \centering
     \caption{Best-fit JAM model and uncertainties (HST/F160W).}
     \footnotesize
    \begin{tabular}{|c|c|cc|}
    \hline\hline
 \small (1) & (2) & (3)& (4)   \\ 
\small {\bf Optimization}& Parameters $\equiv$ Initial Parameters & best-fit & 1$\sigma$   \\  
 \hline
\small   {\bf 1}          & $\log_{10}(M_{\rm BH}/$\Msun) $=7$&  7.12  & $\pm{0.03}$   \\
\small(Constant       &\ml$_{\rm F160W}$ (\Msun/\Lsun) = 3.5  &  3.38  &$\pm{0.15}$ \\
\small\ml)                &     $i (^{\circ})$ = 65                  &  66.5 &$\pm{15.5}$ \\
 \small                      &                  $\beta_r=-0.5$                     &$-$0.32& $\pm{0.06}$  \\  
 \hline 
\small    {\bf 2}       &$\log_{10}(M_{\rm BH}/$\Msun) $=7.12$& 7.11    & $\pm{0.03}$  \\
\small(Varying       & \ml$_{2\farcs0}$ (\Msun/\Lsun) = 3.38   & 3.40   & $\pm{0.02}$  \\
\small\ml)              &                  $i (^{\circ})$ = 66.5                &    65   &  $\pm{13}$   \\
\small                    &                  $\beta_r=-0.32$                     &$-$0.33& $\pm{0.03}$  \\ 
\hline   
\end{tabular}
\label{jamresults_F160W}
\tablefoot{The search ranges for the model parameters were kept fixed as follows: $0 \leq \log_{10}(\Mbh/\Msun) \leq 9$, $-1 \leq \beta_r \leq 1$, $30^\circ \leq i \leq 90^\circ$, and $0 \leq \ml_{\rm F814W}$ or \ml$_{2\farcs0} \leq 20$~\Msun/\Lsun. Columns 3 and 4 present the best-fit parameters and 1$\sigma$ (16–84\%) errors, respectively.} 
\end{table}

\begin{figure*}[!h]    
    \centering
    \includegraphics[width=0.49\textwidth]{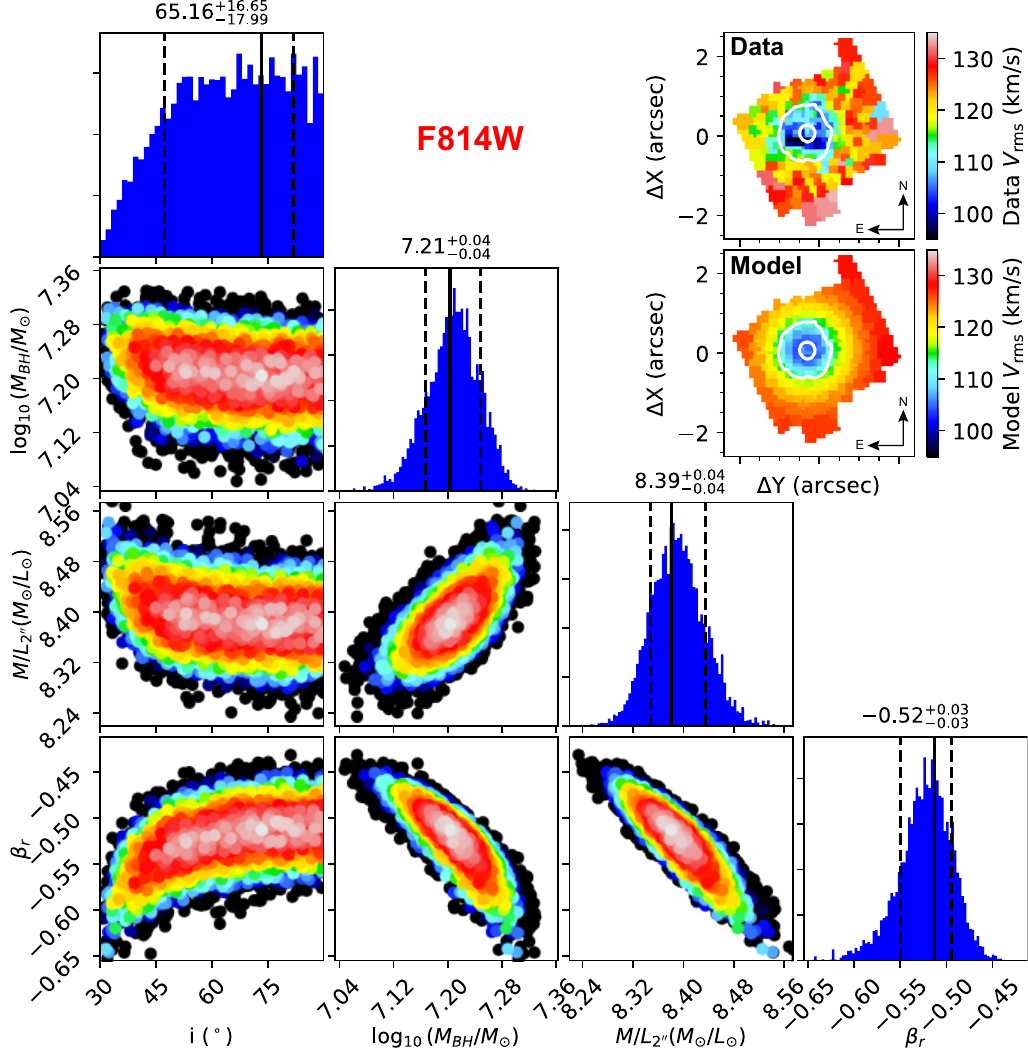} 
    \hspace{2mm}
    \includegraphics[width=0.49\textwidth]{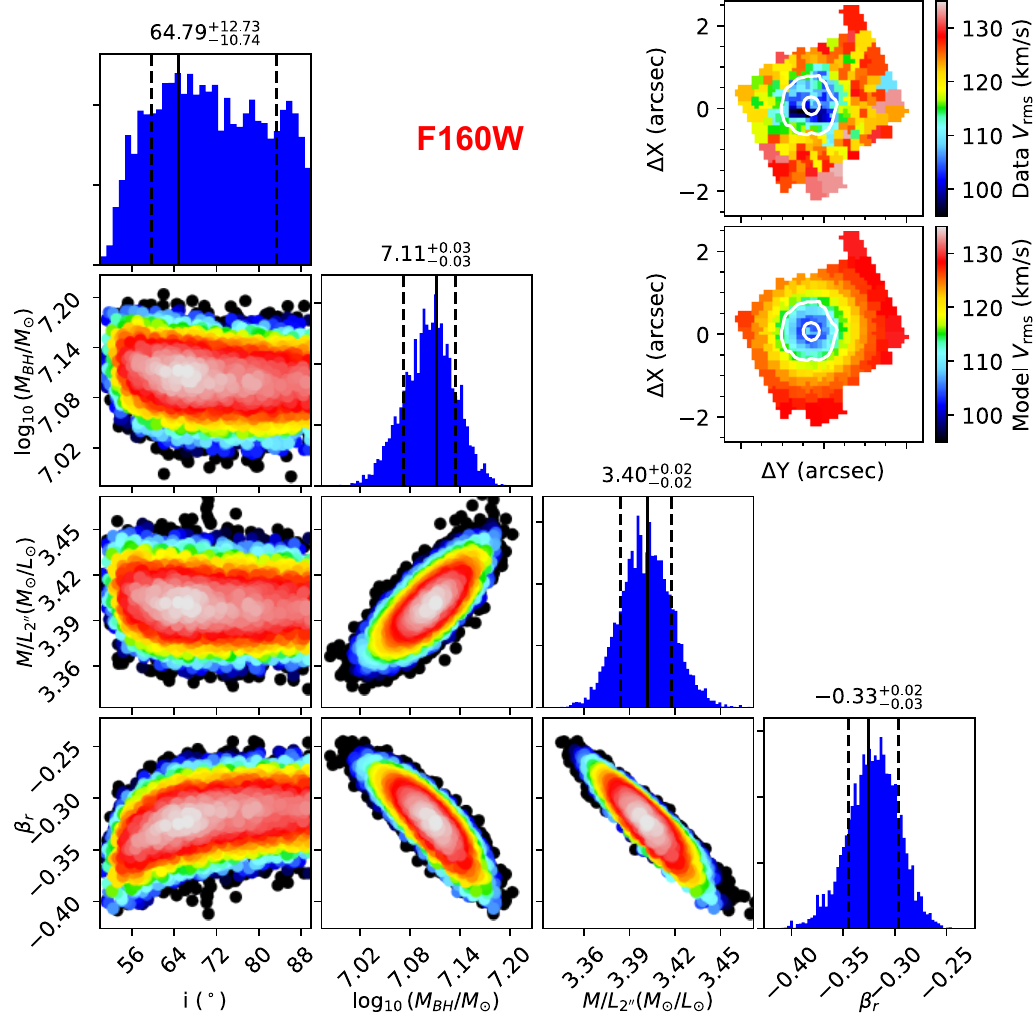} 
     \caption{Posterior distributions for the radial \ml$_{\rm F814W}$ ({left}) and \ml$_{\rm F160W}$ ({right}) variation profiles obtained after the burn-in phase of the \textsc{adamet} MCMC optimization process for the JAM applied to the stellar kinematics of the nucleus of NGC 4736, measured using the NIRSpec G235H/F170LP observations, are presented. In each triangle plot, the analysis involves four parameters ($i$, $M_{\rm BH}$, $M/L_{2\farcs0}$, and $\beta_r$), shown as scatter plots for their 2D projected PDF with white indicates maximum likelihood and a CL at 1$\sigma$, while black corresponds to a confidence level at 3$\sigma$. The top histograms display the 1D projected distributions of the best-fitting model parameters, including the best-fit values (indicated by vertical black lines, corresponding to the highest likelihood within the PDF) and 3$\sigma$ uncertainty (indicated by either side black dash lines). Insert plots in the top-right of each triangle plot illustrate the agreement by directly comparing the $V_{\rm rms}$ data with that generated from the best-fit JAM using the radially varying \ml$_{\rm F814W}$ profile ({left}) and the constant \ml$_{\rm F160W}$ value ({right}) in the same velocity range. It is evident that the constrained model shows excellent agreement with the data across the NIRSpec FoV.}      
    \label{jam_mcmc}
\end{figure*}

The nucleus of NGC 4736 is identified as a typical spiral with kinematics best described using spherical projections. To model its dynamics, we compared the observed $V_{\rm rms}$ map with predictions from JAM with a spherically axisymmetry alignment of the velocity ellipsoid, based on the Jeans equations. This approach aims to estimate \Mbh\ (sampled in logarithmic scale) and constrain additional linearly scaled parameters: orbital anisotropy ($\beta_r$), \ml$_{\rm F814W}$ (or \ml$_{\rm F160W}$), and inclination ($i$). This JAM is well-suited for less rotating spirals \citep{Cappellari08}. It predicts the mean LOSVD, assuming an axially symmetric velocity ellipsoid aligned radially ($\sigma_r \neq \sigma_\theta = \sigma_\phi$). This spherical configuration is specified by setting {\tt align=‘sph’} in the \textsc{jam\_axi\_proj} procedure of the \textsc{JamPy}\footnote{v7.2.0: \url{https://pypi.org/project/jampy/}} package \citep{Cappellari20}. The modelling accounts for the NIRSpec PSF and uses fixed parameter search ranges and initial guesses as detailed in Table \ref{jamresults}.  

We performed a Markov chain Monte Carlo (MCMC) simulation within the JAM framework to explore the best-fit parameter space (\Mbh, \ml$_{\rm F814W\,or\,F160W}$ or \ml$_{2\farcs0}$, $\beta_r$, $i$) and estimate the associated uncertainties. The adaptive Metropolis algorithm \citep{Haario01}, implemented in a Bayesian framework using the \textsc{adamet}\footnote{v2.0.9: \url{https://pypi.org/project/adamet/}} package \citep{Cappellari13a}, was utilized for this purpose.  The MCMC chains comprised $3 \times 10^4$ iterations, with the initial 20\% discarded as a burn-in phase. The remaining 80\% of the iterations were used to construct the full PDF. The best-fit parameters were identified as those corresponding to the highest likelihood within the PDF. Statistical uncertainties for all free parameters were calculated at the 1$\sigma$ (16–84\%) and 3$\sigma$ (0.14–99.86\%) CL (see Fig. \ref{jam_mcmc}).

For each stellar-light MGE model (i.e., either F814W in Table \ref{mgetab} or F160W in Table \ref{mgetab_F160W}),  we conducted two separate MCMC optimization procedures, referred to as Optimization 1 and Optimization 2 (or Opt. 1 and Opt. 2 in Table \ref{jamresults} for the F814W MGE model and in Table \ref{jamresults_F160W} for the F160W MGE model), for the JAM modelings. The first assumes a constant \ml$_{\rm F814W\,or\,F160W}$ profile, while the second incorporates a radially varying \ml$_{\rm F814W\,or\,F160W}$ profile with a linear gradient (Fig. \ref{jam_mcmc} for its best-fit parameters constrained through MCMC). 

The dynamical results presented in Sect. \ref{bh} for the F814W MGE model were derived using our adopted LOSVD obtained from the empirical XSL and the stellar mass listed in Table \ref{mgetab}. To assess the potential impact of template choice on the SMBH mass measurement uncertainty for NGC 4736, we repeated a MCMC optimization using the LOSVD derived from the PHOENIX synthetic stellar spectra library (Appendix \ref{phoenix} and Fig. \ref{jwst_PHOENIX_maps}). This test yielded a totally consistent JAM model with \Mbh\ $=(1.55 \pm 0.20) \times 10^7$ \Msun, \ml$_{2\farcs0} = 8.41 \pm 0.04$ \Msun/\Lsun, orbital anisotropy $\beta_r = -0.50 \pm 0.03$, and an inclination angle of $i = 64^\circ \pm 18^\circ$ for the F814W mass model and  \Mbh\ $=(1.28 \pm 0.20) \times 10^7$ \Msun, \ml$_{2\farcs0} = 3.41 \pm 0.02$ \Msun/\Lsun, orbital anisotropy $\beta_r = -0.35 \pm 0.03$, and an inclination angle of $i = 64.8^\circ \pm 18.3^\circ$ for the F160W mass model, suggesting the systemic related to the choice of stellar spectra libraries is minimal.

\end{appendix}

\end{document}